\documentclass[aps,pra,superscriptaddress,nofootinbib,twocolumn,showpacs,10pt]{revtex4-1} 

\usepackage{amsmath,amssymb,graphicx}
\usepackage{todonotes}
\usepackage{subfigure}
\usepackage[export]{adjustbox}
\usepackage{array}
\usepackage[english]{babel}
\usepackage{mathtools}
\DeclarePairedDelimiter{\ceil}{\lceil}{\rceil}
\newcommand{\etal}{{\it et al.}}

\usepackage[normalem]{ulem}

\newcommand{\gib}{\gamma_\mathrm{ib}}
\newcommand{\gen}{\gamma_n^e}
\newcommand{\ghn}{\gamma_n^h}
\newcommand{\gee}{\gamma_e^e}
\newcommand{\Eg}{\mathcal{E}_g}

\newcommand{\Ekin}{\mathcal{E}_\mathrm{k}}
\newcommand{\Ec}{\mathcal{E}_c}
\newcommand{\sigmol}{\sigma_{\mathrm{mol}}}
\newcommand{\rhomol}{\rho_{\mathrm{mol}}}

\newcommand{\rhoii}{\rho_{\mathrm{ii}}}
\newcommand{\rhon}{\rho_n}
\newcommand{\nufi}{\nu_{\mathrm{fi}}}
\newcommand{\tmre}{t_{\mathrm{MRE}}}
\newcommand{\tdre}{t_{\mathrm{DRE}}}
\newcommand{\Fav}{F_{\mathrm{av}}}
\newcommand{\Fth}{F_{\mathrm{th}}}

\begin{document}
\title{Delayed-rate equations model for femtosecond\\ laser-induced breakdown in dielectrics}
\author{Jean-Luc D\'eziel}
\affiliation{D\'epartement de physique, de g\'enie physique et d'optique, Universit\'e Laval, Qu\'ebec G1V 0A6, Canada}
\author{Louis J. Dub\'e}
\email{Louis.Dube@phy.ulaval.ca}
\affiliation{D\'epartement de physique, de g\'enie physique et d'optique, Universit\'e Laval, Qu\'ebec G1V 0A6, Canada}

\author{Charles Varin}
\email{charles.varin@cegepoutaouais.qc.ca}
\affiliation{D\'epartement de physique, de g\'enie physique et d'optique, Universit\'e Laval, Qu\'ebec G1V 0A6, Canada}
\affiliation{C\'egep de l'Outaouais, Gatineau, Qu\'ebec J8Y 6M4, Canada}

\begin{abstract} 
Experimental and theoretical studies of laser-induced breakdown in dielectrics provide conflicting conclusions about the possibility to trigger ionization avalanche on the sub-picosecond time scale and the relative importance of carrier-impact ionization over field ionization. On the one hand, current models based on single ionization-rate equations do not account for the gradual heating of the charge carriers which, for short laser pulses, might not be sufficient to start an avalanche. On the other hand, models based on multiple rate equations that track the carriers kinetics rely on several free parameters, which limits the physical insight that we can gain from them. In this paper, we develop a model that overcomes these issues by tracking both the plasma density and carriers' mean kinetic energy as a function of time, forming a set of delayed rate equations that we use to match the laser-induced damage threshold of several dielectric materials. In particular, we show that this simplified model reproduces the predictions from the multiple rate equations, with a limited number of free parameters determined unambiguously by fitting experimental data. A side benefit of the delayed rate equations model is its computational efficiency, opening the possibility for large-scale, three-dimensional modelling of laser-induced breakdown of transparent media.
\end{abstract}


\maketitle


\section{Introduction}

Computer modelling of strong-field optical phenomena in dielectrics driven by intense laser radiation is essential to understand the fundamental processes in play, e.g., during laser micro-machining, laser surgery, and high-harmonic generation in solids, to name a few. Mechanisms for laser-induced breakdown were identified and studied in various contexts \cite{gallais2015,jing2012,chimier2011,christensen2009,jupe2009,jia2006,rethfeld2004,kaiser2000}. In the accepted picture, plasma formation in laser-driven dielectrics proceeds as follows. (1) Charge carriers are first created by field ionization (FI). (2) The charge carriers absorb energy from the laser field \textit{via} inverse bremsstrahlung heating (IBH). (3) The hot charge carriers create new, cold ones through carrier-impact ionization (II). (4) The carriers created by II, in turn, gain energy from the laser field and create new carriers by II, and so on. This multiplication of charge carriers \textit{via} II leads to an exponential growth of the plasma density, often referred to as an \textit{ionization avalanche}. This picture applies well when the FI-II interplay has enough time to unfold, e.g., when the pulse duration is in the picosecond range or above. However, current experimental and theoretical studies of laser-induced breakdown in dielectrics provide conflicting conclusions about the relative importance of II over FI and the possibility to trigger ionization avalanche on the sub-picosecond time scale~\cite{balling2013}.  

For example, a pumb-probe experiment in fused silica~\cite{lebugle2014} has shown that a significant amount of ionization can take place after the pump pulse, which cannot be described by FI alone and suggests a delayed II avalanche triggered by slowly-decaying hot plasmon excitations. In contrast, in another experiment in sapphire~\cite{guizard2010}, there was no evidence of ionization avalanche. On the theory side, calculations based upon a Fokker-Planck equation in \cite{stuart1995} lead to a strong dominance of II over FI while the simulations in \cite{shcheblanov2012} predict kinetic energies of the charge carriers that are too low for II to be significant. It was also suggested that the condition to trigger an avalanche should be given by the laser fluence instead of the pulse duration, but the predicted threshold values differ by more than an order of magnitude (see, e.g.,~\cite{rethfeld2006,petrov2008}). 

Actually, experiments involve different materials and laser parameters, which makes a direct comparison between them difficult. Other challenges lie in the theoretical models that are currently used to interpret the experimental observations. On the one hand, current models based on single ionization-rate equations (SRE) do not account for the gradual heating of the charge carriers which, for short laser pulses, might not be sufficient to start an avalanche. On the other hand, models based on multiple rate equations (MRE) that track the carrier kinetics on discrete energy levels rely on several free parameters, which limits the physical insight that we can gain from them. While calculation of the FI rates with the Keldysh theory~\cite{keldysh1965} is well established, models for IBH and II can vary significantly. For example in refs.~\cite{rethfeld2004,christensen2009,bourgeade2010,gallais2015}, plasma formation was modelled within similar theoretical frameworks, but assumed different IBH rates, thus influencing directly the efficiency of II and leading to conflicting conclusions about the relative importance of II over FI and the occurrence of ionization avalanche in short pulses.

In this paper, we describe a model that overcomes the issues associated with the SRE and MRE models by tracking both the plasma density and carriers' mean kinetic energy as a function of time, forming a set of delayed rate equations (DRE) that we use to match the laser-induced damage threshold of several dielectric materials. In particular, we show that this simplified model reproduces the predictions from the multiple rate equations, with a limited number of free parameters determined unambiguously by fitting experimental data. A side benefit of the DRE model is its computational efficiency, opening the possibilities for large-scale, three-dimensional modelling of laser-induced breakdown of transparent media.

The paper is organized as follows. First in Sec.~\ref{eq:model_overview} we present an overview of the single- and multiple-rate models. Next in Sec.~\ref{sec:delayed_rate_model}, we describe the proposed delayed-rate equations model in details. The three models (SRE, MRE, and DRE) are compared in Sec.~\ref{sec:comparison}. In Sec.~\ref{sec:dre_and_experiments}, we show how the DRE model can fit experimental data for the damage threshold in several dielectric materials. In Sec.~\ref{sec:discussion}, we discuss some of the limitations of the DRE model and, ultimately, we conclude in Sec.~\ref{sec:conclusion}. Three Appendices gather some of the technical aspects of the model and of its implementation. All calculations are performed using a Python package that we made available online \cite{pyplasma}.

\section{Overview of current rate equation models for laser-induced plasma formation in dielectrics}\label{eq:model_overview}

The modelling of the laser-induced polarization and breakdown dynamics in solid-state dielectrics is typically composed of three complementary pieces. (1) A model for the polarization density from bound electrons. (2)~A model for the evolution of the conduction band population due to field ionization, impact ionization, and electron-hole recombination. (3)~A model for the free-current density associated with the charge carriers (electrons and holes). This approach provides great modelling flexibility and a fair description of the underlying physics on a cycle-averaged statistical level (see Sec.~\ref{sec:discussion} for discussion). Below, we provide an overview of two established population-dynamics models. For reviews of bound and free currents models see, e.g., refs.~\cite{Couairon2011,balling2013,Kolesik2014a,varin2018}.

\subsection{Single rate equation} \label{sec:sre}

The simplest way to describe plasma formation in dielectrics while accounting for both FI and II is the single rate equation (SRE)~\cite{stuart1995}:
\begin{align}\label{eq:sre} 
\frac{\mathrm{d}\rho}{\mathrm{d}t} = \underbrace{\nu_\mathrm{fi}\rhon}_{\mathrm{FI}} + \underbrace{\alpha\rho I \rhon}_{\mathrm{II}} -\underbrace{ \gamma_r\rho}_{\mathrm{RE}}, 
\end{align}
where $\rho$ represents the carrier density. The first two terms on the right hand side ---associated with the field ionization (FI) rate $\nufi$ and the impact ionization (II) rate $\alpha\rho I$, respectively ---weighted by the density of neutral molecules or atoms $\rhon=(\rhomol-\rho)$ (if we account for single ionization at most). Here, $\rhomol$ is the molecular density, $\alpha$ is the impact rate coefficient, and $I=c\epsilon_0n_0E^{2}/2$ is the cycle-averaged laser intensity, with $n_0$ being the linear refractive index. The last term accounts for the recombination (RE) of electrons and holes at a rate $\gamma_r$.

The SRE model was first developed upon empirical observations. The linear relation between the II rate and intensity can be justified by the linearity between the heating rate of the charge carriers and the laser intensity [see Eq.~\eqref{eq:gib} below]. As such, the rate at which the electrons and holes gain energy via IBH and the rate at which they give it back \textit{via} II both scale linearly with intensity. However, the linear scaling between II and $\rho$ in Eq.~\eqref{eq:sre} implies that all charge carriers can contribute to II, regardless of their energy. This causes an overestimation of II, especially at low fluence. In fact, an electron or a hole needs to acquire a minimum energy $\Ec$ to allow a collision where a new valence electron crosses the band gap and reaches the conduction band. To respect both energy and momentum conservation, the critical kinetic energy required for II to be possible is \cite{kaiser2000}
\begin{align}\label{eq:E_critical} 
\Ec = \left( 1+\frac{m_r}{m_h} \right)(\mathcal{E}_g+\mathcal{E}_p),
\end{align}
where $m_r$ is the reduced mass, $m_h$ is the effective hole-mass, $\mathcal{E}_g$ is the bandgap energy, and $\mathcal{E}_p$ is the ponderomotive energy. See Appendix~\ref{appendix:masses} for the definitions associated with the mass symbols used, and Appendix~\ref{sec:laser_currents} for the definition of $\mathcal{E}_p$.

The main difference between SRE and more advanced approaches lies in the relation between the II rate and the plasma density $\rho$. In particular, the relation should account for the gradual heating of the carriers by the laser and respect the necessity for them to reach the critical energy $\mathcal{E}_c$ for II to occur. Both the MRE [see Sec.~\ref{sec:mre}] and the DRE [see Sec.~\ref{sec:delayed_rate_model}] models address this issue, although with somewhat different ingredients.

\subsection{Multiple rate equations} \label{sec:mre}

To gain insight into the dielectric breakdown process as a whole, Kaiser \etal ~\cite{kaiser2000} have developed a first-principle model that accounts for the various interactions between light, phonons, and the charge carriers. It describes how FI stacks electrons in a single energy level at the bottom of the conduction band (CB), creating a sharp spike in the energy distribution at $\hbar\omega$. When subsequent photon absorption takes place, a new spike appears at $2\hbar\omega$, then another one at $3\hbar\omega$, and so on. After a few femtoseconds, these spikes broaden and disappear due to collisions (thermalization). By tracking dynamically the energy distribution, the number of charge carriers having a minimum kinetic energy of $\Ec$ [see Eq.~\eqref{eq:E_critical}] to contribute to II is then known. II rates can then be scaled with respect to this reduced population (carriers with $\mathcal{E} > \mathcal{E}_c$) instead of the entire distribution as done in the SRE model. 

A drawback of Kaiser \etal's ~\cite{kaiser2000} approach is the large number of coupled differential equations that need to be solved (a few hundreds in the case of fused silica). To find a middle ground between simplicity (SRE) and completeness (Kaiser \etal~\cite{kaiser2000}), Rethfeld has developed a multiple rate equations model (MRE) by neglecting thermalization~\cite{rethfeld2004}. By doing so, the \emph{spikiness} of the energy distribution is fully preserved. The energy distribution can then easily be discretized in $k=\ceil{\Ec/\hbar\omega}$ energy levels (plus one for the zeroth level), each separated by increments of $\hbar\omega$ and associated with an individual population. A rate equation for each level is then solved to track the entire energy distribution. For the electrons in the CB, these rate equations are
\begin{subequations}
	\begin{align} 
	\frac{\mathrm{d}\rho_0}{\mathrm{d}t} &= \nu_\mathrm{fi}\rhon - \gib^{e}\rho_0 +2\gamma^e_n\rho_k +\gamma^h_n\rho_k^h - \gamma_r\rho_0, \label{eq:rho0} \\[-2mm]
	&~~\vdots \nonumber \\[-2mm]
	\frac{\mathrm{d}\rho_j}{\mathrm{d}t} &= \gib^{e} (\rho_{j-1}-\rho_j) - \gamma_r \rho_j \quad ; \quad 1\leq j < k, \label{eq:rhoj} \\[-2mm]
	&~~\vdots \nonumber \\[-2mm]
	\frac{\mathrm{d}\rho_k}{\mathrm{d}t} &= \gib^{e}\rho_{k-1} -\gamma^e_n\rho_k -\gamma_r \rho_k, \label{eq:rhok} \\
	\rho &= \sum_{j=0}^k \rho_j.
	\end{align} \label{eq:mre}
\end{subequations}
The population at the zeroth level $\rho_0$ (the bottom of the CB) is seeded by FI \textit{via} the first term of the right hand side of Eq. \eqref{eq:rho0}. The next term represents the electrons that are removed from the zeroth level as they absorb photons at a rate $\gib^e$. Each time an electron is removed from any of the $(j-1)$th energy level because of IBH absorption, it is added to the $j$th level, as described by Eq.~\eqref{eq:rhoj}. After $k$ subsequent photon absorptions, electrons reach the upper $k$th level, with a kinetic energy of at least $\Ec$. At this level, IBH is artificially stopped to limit the number of rate equations. From then on, electrons can collide with neutral molecules and cause II events at a rate $\gen$. These electrons then lose their kinetic energy and fall back to the zeroth level while bringing a second electron from the valence band (VB) to the CB (the zeroth level) [see the third term of Eq. \eqref{eq:rho0}]. Plasma relaxation at the rate $\gamma_r$ is also included across all energy levels. 

We have extended the original model of Rethfeld ~\cite{rethfeld2004} to account for II events caused by holes. To do so, the term $\ghn\rho_k^h$ was added to Eq. \eqref{eq:rho0} with the hole-neutral molecule collision rate $\ghn$ and the population of holes in the $k$th level $\rho_k^h$. The latter is calculated with a second set of $k+1$ rate equations that tracks the holes energy distribution. This second set is similar to Eqs. \eqref{eq:mre}, but with $\rho^h_j$, $\ghn$ and $\gib^h$ instead of $\rho_j$, $\gen$ and $\gib^e$ respectively. In the special case where electrons and holes have the same mass $m_e=m_h$, both sets of $k+1$ equations are equivalent and only one has to be solved, with $\ghn=\gen$, $\gib^{h}=\gib^{e}$ and $\rho_j^{h}=\rho_j$.

Summing Eqs.~\eqref{eq:mre} leads to the global plasma formation rate as:
\begin{align}
\frac{\mathrm{d}\rho}{\mathrm{d}t} &= \nu_\mathrm{fi}\rhon + \sum_{s=e,h}\gamma^s_n\rho_k^s - \gamma_r\rho, \label{eq:mre2}
\end{align}
where $s=\{e,h\}$ stands for electrons and holes, respectively. Eq.~\eqref{eq:mre2} (MRE) and Eq.~\eqref{eq:sre} (SRE) are identical, except for the second term, associated with II. For the MRE model, the II rate scales with the upper-level populations $\rho_k^e$ and $\rho_k^h$. Since the $k$th energy level is populated only after $k$ subsequent photon absorption, II is effectively delayed with respect to FI. The delay for II to unfold is approximately
\begin{align}
\tmre &= \left[( \sqrt[k]{2}-1) \gib^{e}\right]^{-1}, \label{eq:tmre} \\
\tmre &\to  \frac{1}{\gib^e}\frac{k}{\ln 2} ~~\mathrm{for}~~ k\gg 1,
\end{align}
when accounting only for II events caused by CB electrons. To account also for VB holes, a distinct delay is set by replacing $\gib^{e}$ in Eq. \eqref{eq:tmre} by $\gib^{h}$.

Rethfeld ~\cite{rethfeld2004} has concluded that for pulse duration shorter than $\tmre$, II is negligible compared with FI because charge carriers are not heated enough to reach $\Ec$ before the end of the laser pulse. However IBH is proportional to the laser intensity, i.e., $\gib \propto I$ (see Appendix~\ref{sec:laser_currents}), which suggests that fast, sub-ps carrier heating is possible if the laser intensity is sufficiently high. Thus, a more general condition to trigger an avalanche of ionization through II is $F > I\tmre$~\cite{rethfeld2006}, where $F$ is the laser fluence [see Eq.~\eqref{eq:fluence} for definition].

\section{The delayed-rate equation model}\label{sec:delayed_rate_model}

We describe next a delayed-rate equation (DRE) model which addresses the lack of carrier dynamics of the SRE model, while being simpler and less computationally demanding than MRE. Numerical comparison between DRE, SRE, MRE, and experimental data will follow in Secs.~\ref{sec:comparison} and~\ref{sec:dre_and_experiments}.

We recall that the early energy distribution of the electrons calculated by the full kinetic approach (see ref.~\cite{kaiser2000}) exhibits sharp spikes. This has motivated the development of the MRE model that tracks the electron heating dynamics over discrete momentum levels ($\hbar\omega, 2\hbar\omega, \ldots$). However, these spikes quickly broaden and disappear after only a few femtoseconds, due to collisions that drive the energy distribution towards thermal equilibrium. Following a different strategy than for MRE, that assumes that no thermalization takes place, we rely next on the approximation that on a few-laser-cycle timescale, the thermalization process can be considered as almost instantaneous (See Sec.~\ref{sec:comparison} for a comparison between both approaches). 

\begin{figure}
\centering\vspace{2mm}
\includegraphics[width=1.0\columnwidth]{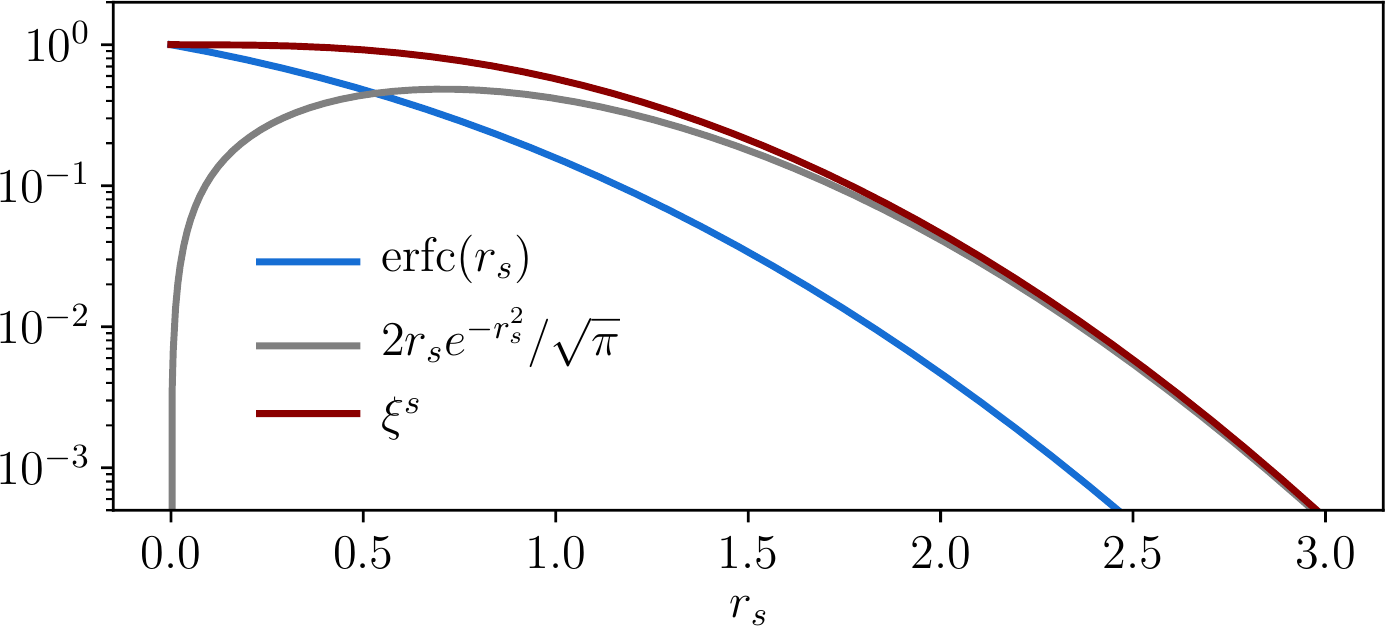}
\caption{(Color online) The fraction $\xi^s$ of the charge carriers that have a larger kinetic energy than $\Ec$ as a function of the dimensionless parameter $r_s$ [see Eq. \eqref{eq:xi}]. The two contributing terms of $\xi^{s}$ are shown for comparison.}\label{figxi}
\end{figure}

Assuming a Maxwellian thermal-equilibrium energy distribution, the fraction of electrons ($s=e$) or holes ($s=h$) that have an energy higher than the critical energy $\Ec$ can be calculated analytically as
\begin{align}\label{eq:xi}
\xi^{s} &= \frac{\int_{\Ec}^{\infty} \mathcal{E}^{1/2} \exp[-3\mathcal{E}/2\Ekin^{s}] \mathrm{d}\mathcal{E}}{\int_0^{\infty} \mathcal{E}^{1/2} \exp[-3\mathcal{E}/2\Ekin^{s}] \mathrm{d}\mathcal{E}} \\
&= \mathrm{erfc}(r_s) + \frac{2r_s}{\sqrt{\pi}} \exp (-r_s^{2}),
\end{align}
where $r_s=\sqrt{3\Ec/2\Ekin^{s}}$. The ratio $\xi^{s}$ and its two contributing terms are shown in Fig.~\ref{figxi} as a function of the dimensionless parameter $r_s$. Notice that the individual contributions are almost equal at $r_s = 0.5$, whereupon the second term rapidly becomes dominant for $r_s > 1$. With $\xi^{s}$, the equation for the charge-carrier density can be written as
\begin{align}\label{eq:dre}
\frac{\mathrm{d}\rho}{\mathrm{d}t} &= \nu_\mathrm{fi}\rhon + \sum_{s=e,h}\gamma^s_n\xi^s\rho - \gamma_r\rho,
\end{align}
which is similar to the MRE equation~\eqref{eq:mre2}, with  $\rho_k^{s}$ replaced by $\xi^{s}\rho$. The other terms, associated with field ionization ($\nu_\mathrm{fi}\rhon$) and electron-hole recombination ($\gamma_r\rho$) are identical to both the SRE and MRE models [see Eqs.~\eqref{eq:sre} and~\eqref{eq:mre2}, respectively].

The simplicity of the DRE model comes from the possibility to track the mean kinetic energy $\Ekin^{s}$ of the electrons and holes, instead of the multiple level populations of MRE [see Eqs.~\eqref{eq:mre}]. This is done with the single ordinary differential equation that follows:
\begin{align}\label{eq:Ek}
\frac{\mathrm{d}\Ekin^s}{\mathrm{d}t} &= \gib^s\hbar\omega - \gamma^s_n\xi^s\Ec - \Ekin^{s}\left[\nu_\mathrm{fi}\frac{\rhon}{\rho}+\sum_{s=e,h}\gamma^s_n\xi^s\right].
\end{align} 
The first term on the right hand side is associated with photon absorption through IBH. The second term represents the kinetic energy lost in an II event. The final terms (in square brackets) ensure energy conservation for each ionization event and redistribute the kinetic energy among the new charge carriers generated by FI and II. 

Some insight into the general behaviour of the DRE model would be useful before undertaking the numerical comparison with SRE and MRE in Sec.~\ref{sec:comparison}. Initially, i.e., at time $t = 0$, there are no charge carriers and the ratio $\xi^{e}$ is identically zero. Field ionization will then bring electrons to the conduction band, and these electrons will be gradually heated up by the laser field, thus increasing the average electron kinetic energy and the value of $\xi^{e}$. At some point, we can expect that laser heating will be balanced by the energy losses from II, i.e., $\gib^{e}\hbar\omega \simeq \gamma_n^{e}\xi^{e}\Ec$. If furthermore, we neglect recombination ($\gamma_r=0$) and the depletion of the valence electron population ($\rhon\simeq\rhomol$), the electron population in the conduction band is roughly given by:
\begin{align}\label{eq:dre_approx}
\frac{\mathrm{d}\rho^e}{\mathrm{d}t} &\simeq \nu_\mathrm{fi}\rhomol + \left(\frac{\gib^{e}\hbar\omega}{\mathcal{E}_c}\right)\rho^e,
\end{align}
whose solution is
\begin{align}\label{eq:rhot}
\rho^e(t) = \nufi\rhomol\frac{\Ec}{\gib^e\hbar\omega} \left[ \exp \left( \frac{\gib^e\hbar\omega}{\Ec}t \right) - 1 \right]. 
\end{align}
Note that $\rho^e(0)=0$. Eq.~\eqref{eq:rhot} shows an exponential increase of the free-electron density, a characteristics of an ionization avalanche. The argument in the exponential function gives the following characteristic time
\begin{align}
t_\mathrm{DRE} = \frac{1}{\gib^{e}}\frac{\Ec}{\hbar\omega}.\label{eq:tdre}
\end{align}
In the next section, we will see that Eq.~\eqref{eq:tdre} predicts an avalanche delay that is comparable to that obtained with the MRE model [see Eq.~\eqref{eq:tmre}, with $k\sim \Ec/\hbar\omega$], which suggests that the plasma thermalization dynamics (as described by Kaiser \etal ~\cite{kaiser2000}) has a limited impact on the avalanche process as a whole. 

\section{Numerical analysis of the rate models} \label{sec:comparison}

So far, we have described three rate-equation models (SRE, MRE, and DRE) that track the temporal evolution of the charge-carrier density on a field-cycle-average, statistical level during laser-induced breakdown. We have seen that these models differ only in the way they account for impact ionization and, in particular, for the delay associated with the laser-heating process [compare Eqs.~\eqref{eq:sre},~\eqref{eq:mre2}, and~\eqref{eq:dre}].

We examine the behaviour of the three rate models with respect to impact ionization by computing the ratio of the plasma density generated by impact ionization $\rho_\mathrm{ii}$ over the total plasma density $\rho$ when an harmonic electric field $\tilde{E}(t)=E\cos(\omega t)$ with a constant amplitude $E$ is applied. For each model, the laser intensity $I=cn_0\epsilon_0E^{2}$ is set to obtain $\rho_\mathrm{ii}/\rho\simeq 0.5$ after $t=100~\mathrm{fs}$ ($n_0$ is the refractive index of the dielectric without ionization). This intensity marks, for each model, the turning point where impact ionization becomes dominant ($\rho_\mathrm{ii}/\rho > 0.5$). For the tests that follow, we thus define the fluence threshold for impact ionization avalanche as $F_{\mathrm{av}} = I\cdot 100\,\mathrm{fs}$.

To describe field ionization (FI), we used the Keldysh theory~\cite{keldysh1965} that accounts for both multiphoton and tunnel ionization in a unified framework. To calculate the rate $\nufi$ for solid state materials, we rely on the formalism presented in ref.~\citep{balling2013}. For convenience, we reproduce these equations in Appendix~\ref{sec:fi}, with a slightly different notation. For simplicity, we first neglect recombination ($\gamma_r = 0$) (this contribution will be taken into account later when we compare DRE with experimental data).

In presenting the model equations in Secs.~\ref{sec:sre}, ~\ref{sec:mre}, and~\ref{sec:delayed_rate_model}, an explicit description of the laser heating rate $\gib^s$ and the free-carrier-to-neutral impact rate $\gamma^s_n$ was not given. These two quantities depend on the dynamic properties of the electron-hole plasma. Assuming an harmonic laser electric field $\tilde{E}(t)=E\cos(\omega t)$, the classical Drude model leads to the following expression for the laser-heating rate [see Appendix~\ref{sec:laser_currents}, in particular, Eq.~\eqref{eq:gib}]
\begin{align} \label{eq:gib2}
\gib^s = \frac{\gamma}{\hbar\omega}\frac{q^2 E^2}{2m_s(\gamma^2 + \omega^2)}.
\end{align}
In Eq.~\eqref{eq:gib2}, the plasma damping parameter $\gamma$ accounts effectively for collisions between free carriers (e.g., $\gamma_e^e$, $\gamma_e^h$, $\gamma_h^h$, $\ldots$) and phonons. For direct collisions between charge carriers and neutral molecules $\gamma^s_n$, we used the model of ref.~\cite{balling2013}, i.e., 
\begin{align} 
\gamma^s_n = \sigmol \rhon \sqrt{\frac{2\Ekin^{s}}{m_s}}, \label{eq:gn}
\end{align}
where $\sigmol$ is the molecular impact cross-section. However, the results obtained with DRE are nearly unaffected whether we use Eq.~\eqref{eq:gn} or a constant value for $\gamma_n^{s}$. This observation is supported by the work reported in ref.~\cite{rethfeld2004}, where it is shown that the value of $\gamma_n^s$ (or that given by the underlying model) has a small influence, as long as $\gamma_n^s \gg \gib^s$. 

\begin{figure}
	\centering\vspace{2mm}
	\includegraphics[width=1.0\columnwidth]{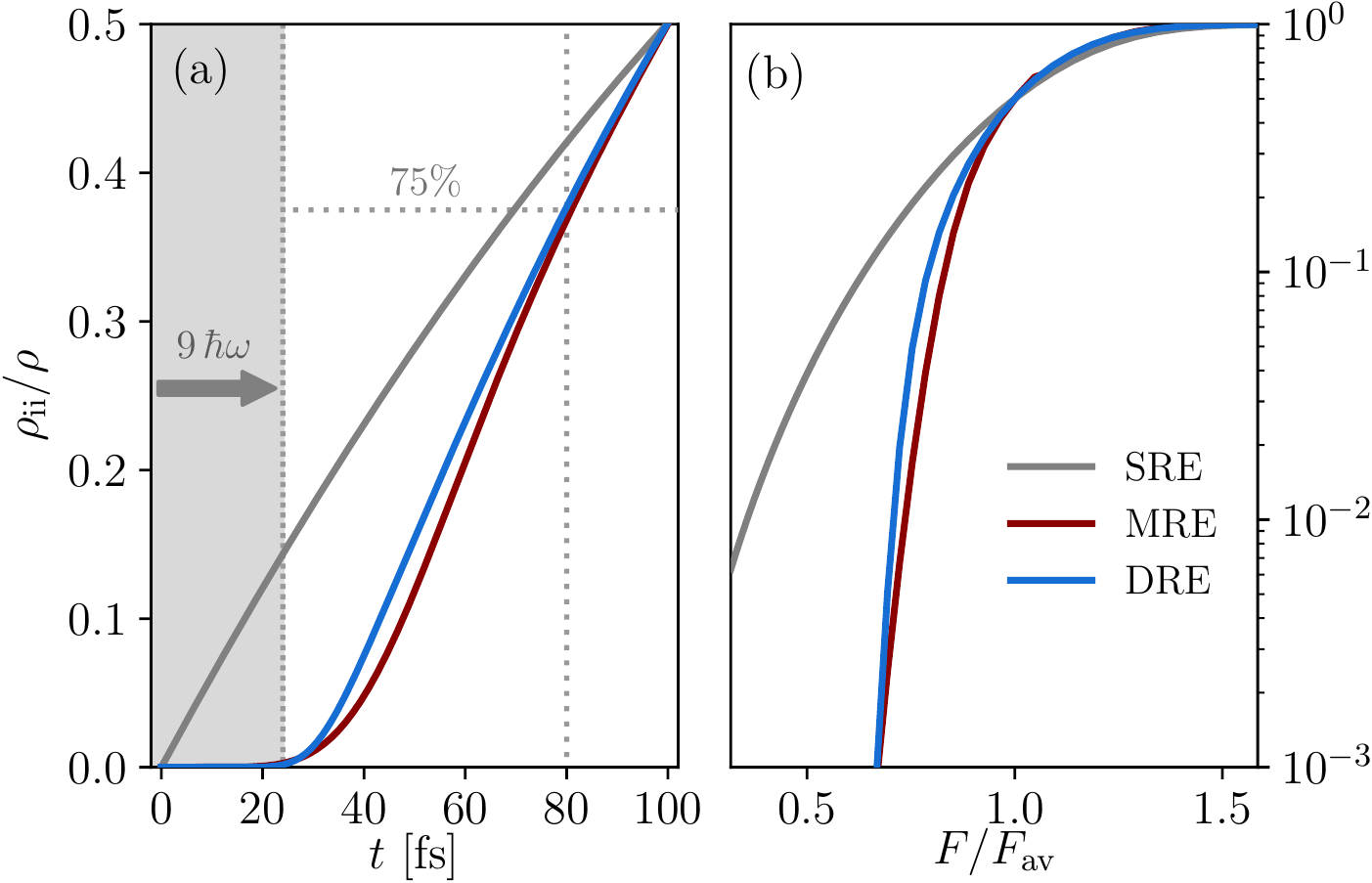}
	\caption{(Color online) Relative contribution of impact ionization ($\rhoii$) over the global ionization yield ($\rho$) obtained with the SRE, MRE, and DRE models. In (a), impact ionization in SRE starts immediately, while MRE and DRE show a 9-photon absorption delay needed for the first charge carriers to be heated above the critical energy $\mathcal{E}_c$. The fluences needed to reach the avalanche condition $\rho_{\mathrm{ii}}/\rho = 0.5$ within $100\,\mathrm{fs}$ are $\Fav^\mathrm{SRE}=0.420\,\mathrm{J/cm}^2$, $\Fav^\mathrm{MRE}=0.459\,\mathrm{J/cm}^2$, and $\Fav^\mathrm{DRE}=0.356\,\mathrm{J/cm}^2$. In (b), the contribution from impact ionization drops quickly when the laser fluence $F$ is below the avalanche threshold $\Fav$. The drop is more pronounced for MRE and DRE. For each model, the fluence $F$ is normalized by the respective $\Fav$ value. Model parameters are $\lambda=800\,\mathrm{nm}$, $\Eg=9\,\mathrm{eV}$, $m_e=m_h=m_0$, $\rhomol=2\times 10^{28}\,\mathrm{m}^{-3}$, $\sigmol=10^{-19}\,\mathrm{m}^2$, $n_0=1.5$, $\gamma_r=0$, $\gamma=1\,\mathrm{fs}^{-1}$ and $\alpha=4\,\mathrm{cm}^2/\mathrm{J}$.}\label{figRatio}
\end{figure}

Numerical results for a fictitious material whose properties are comparable to SiO$_2$ are presented in Fig.~\ref{figRatio}. As expected, we see in Fig.~\ref{figRatio}(a) that the density of charge carriers generated via II predicted by MRE and DRE are delayed with respect to SRE. For both MRE and DRE, the delays follow the predicted values from Eqs.~\eqref{eq:tmre} and~\eqref{eq:tdre}, $\tmre=83.2$~fs and $\tdre=76.5$~fs, respectively. As shown in Fig.~\ref{figRatio}(b), for MRE and DRE the contribution from II to the total plasma density drops sharply for fluence below the avalanche threshold ($F<\Fav$). Above threshold ($F>\Fav$), all three models show a similar trend.

We recall that MRE has been developed in the limit of an infinite thermalization time, whereas DRE was developed in the limit of an infinitesimal thermalization time. Our numerous tests reveal that the plasma formation rates are quite similar in both limits, providing compelling evidence that the thermalization time has a small impact upon the plasma formation process as a whole.

To get more insight into the DRE model, we have considered a more realistic scenario where a strong laser pulse is incident on a fictitious material similar to SiO$_2$. The electric field envelope of the laser pulse 
in vacuum is modelled by a Gaussian function:
\begin{align} 
E_\mathrm{vac}(t) = E_0 \exp \left[ -2\ln(2)\left( \frac{t}{\tau} \right)^2 \right],
\end{align}
where $\tau$ is the full-width at half-maximum (FWHM) duration of the pulse. The laser intensity and fluence in vacuum are then
\begin{align} 
I_\mathrm{vac}(t) = c\epsilon_0 |E_\mathrm{vac}(t)|^2 = c\epsilon_0 E_0^2 \exp \left[ -4\ln(2)\left( \frac{t}{\tau} \right)^2 \right],
\end{align}
and
\begin{align} \label{eq:fluence}
F = \int_{-\infty}^\infty I_\mathrm{vac}(t) \mathrm{d}t = \frac{c\epsilon_0 E_0^2 \tau}{2} \sqrt{\frac{\pi}{\ln(2)}},
\end{align}
respectively. To account for the intrinsic refractive index $n_0$ of the material, as well as for the laser-induced metalization, we computed the electric field in the bulk with:
\begin{align} 
E_\mathrm{bulk}^{2}(t) = E_\mathrm{vac}^{2}(t) \frac{1-R}{n_0}, \label{eq:Ebulk}
\end{align}
where
\begin{align} 
R = \left| \frac{n-1}{n+1} \right|^{2}
\end{align}
and
\begin{align} 
\label{eq:refractive_index}
n^{2} = n_0^{2} - \frac{\omega_p^{2}}{\omega^{2} + i\omega\gamma}.
\end{align}
This last relation is obtained from the Drude model with a plasma frequency $\omega_p^{2}=q^{2}\rho/\epsilon_0m_r$, updated dynamically as the carrier density $\rho$ grows. 

\begin{figure}
\centering\vspace{2mm}
\includegraphics[width=1.0\columnwidth]{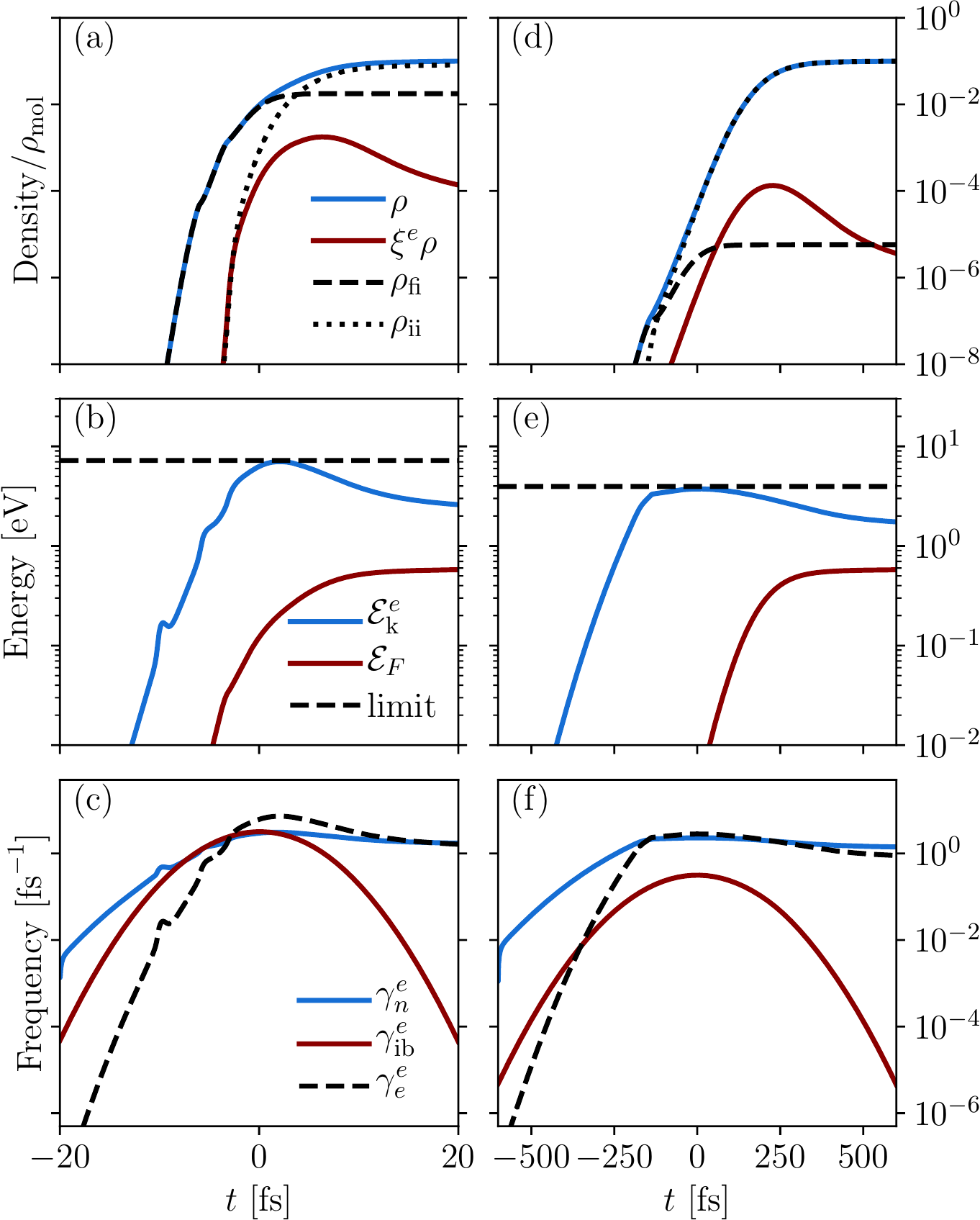}
\caption{(Color online) Two examples of solutions for the DRE. In the left column [(a) to (c)] are shown results for a short pulse duration $\tau=10$ fs and $F=1.6$ J/cm$^{2}$ and in the right column [(d) to (f)] are shown results for $\tau=300$ fs and $F=4.8$ J/cm$^{2}$. In both cases, the fluence is adjusted to reach 10 \% of ionized molecules. Parameters are $\lambda=800$ nm, $\Eg=9$ eV, $m_e=m_h=m_0$, $\rhomol=2\cdot 10^{28}$ m$^{-3}$, $\sigmol=10^{-19}$ m$^2$, $n_0=1.5$, $\gamma_r=0$ and $\gamma=1$ fs$^{-1}$.}\label{figDre}
\end{figure}

By solving DRE with the gaussian laser source $E_\mathrm{bulk}(t)$, we obtain the results displayed in Fig.~\ref{figDre}. In the leading edge of the pulse, most of the plasma comes from FI [Figs.~\ref{figDre}(a) and~\ref{figDre}(d)] as is typically expected. However, as charge carriers get heated up and reach the critical energy $\Ec$, plasma growth switches to II.

We have also compared the average kinetic energy of the electrons $\Ekin^{e}$ to the Fermi energy $\mathcal{E}_F = \hbar^{2}(3\pi^{2}\rho)^{2/3}/2m_e$ [see Figs.~\ref{figDre}(b) and~\ref{figDre}(e)]. Over the entire simulations $\Ekin^{e} > \mathcal{E}_F$, which suggests that using a Fermi-Dirac distribution to get the ratios $\xi^{s}$, instead of a Maxwellian distribution, should not be necessary. This condition is respected in all the calculations performed here.

A rough estimate of an upper limit for the average kinetic energy of the charge carriers $\Ekin^s$ is obtained in the regime where $\gib^{s}\hbar\omega \simeq \gamma_n^{s}\xi^{s}\Ec$ [see also the paragraph before Eq.~\eqref{eq:dre_approx}]. For moderate laser intensity, only a small fraction of carriers effectively reach the critical energy such that $\Ekin^s \ll \Ec$ at all times. In this regime, $r_s \gg 1$ where $\xi^{s}$ is well approximated by $\frac{2r_s}{\sqrt{\pi}} \exp (-r_s^{2})$, and it is then possible, in combination with \eqref{eq:gn}, to obtain an explicit upper bound
\begin{align}
\Ekin^{s} < -\frac{3}{2}\Ec\left[\ln\left(\frac{\gib^{s}\hbar\omega}{2\Ec\sigmol\rhomol}\sqrt{\frac{m_s\pi}{3\Ec}}\right)\right]^{-1}.\label{eq:Emax}
\end{align}
This approximation is in good agreement with the numerical results shown in in Figs.~\ref{figDre}(b) and~\ref{figDre}(e) (see dashed lines). Effectively, Eq.~\eqref{eq:Emax} predicts that the maximum average kinetic energy should not exceed the value given by the right-hand side of the inequality. For example in Fig.~\ref{figDre}(b), the prediction from Eq.~\eqref{eq:Emax} is 7.214 eV and the maximum obtained from the numerical integration of DRE is 7.035 eV (a 2.54\% overestimation). For the longer pulse duration case in Fig.~\ref{figDre}(e), the predicted upper bound is 3.957 eV and the maximal value obtained in the simulation is 3.742 eV (a 5.75\% overestimation).

The laser heating rate $\gib^{e}$ and the electron-neutral collision rate $\gen$ are shown in Figs.~\ref{figDre}(c) and~\ref{figDre}(f). For comparison, we display as well the electron-electron collision rate given by the following formula (see ref.~\cite{christensen2009})
\begin{align}
\gee = \frac{4\pi\epsilon_0}{q^{2}}\sqrt{\frac{6}{m_e}}\left( \frac{2\Ekin^{e}}{3} \right)^{3/2}.
\end{align}
It is then observed that $\gee$ increases rapidly at the leading edge of the pulse, as the plasma gets initially build up by FI. But when II takes over FI, its value levels off to approximately 1 to 10$~\mathrm{fs}^{-1}$, which supports the hypothesis of a fast thermal relaxation and the neglect of the internal thermalization dynamics.

\section{Calibration of the delayed-rate equation model to experiments} \label{sec:dre_and_experiments}

The DRE model presented in Sec.~\ref{sec:delayed_rate_model} depends on a closed set of parameters. Some of them can be directly linked to material properties obtained from experimental measurements or \textit{ab initio} calculations (e.g., $n_0$, the electron-impact cross sections, ...). Below we show how effective values for the remaining parameters can be obtained by fitting the DRE model to damage-threshold data.

\begin{figure}
\centering\vspace{2mm}
\includegraphics[width=1.0\columnwidth]{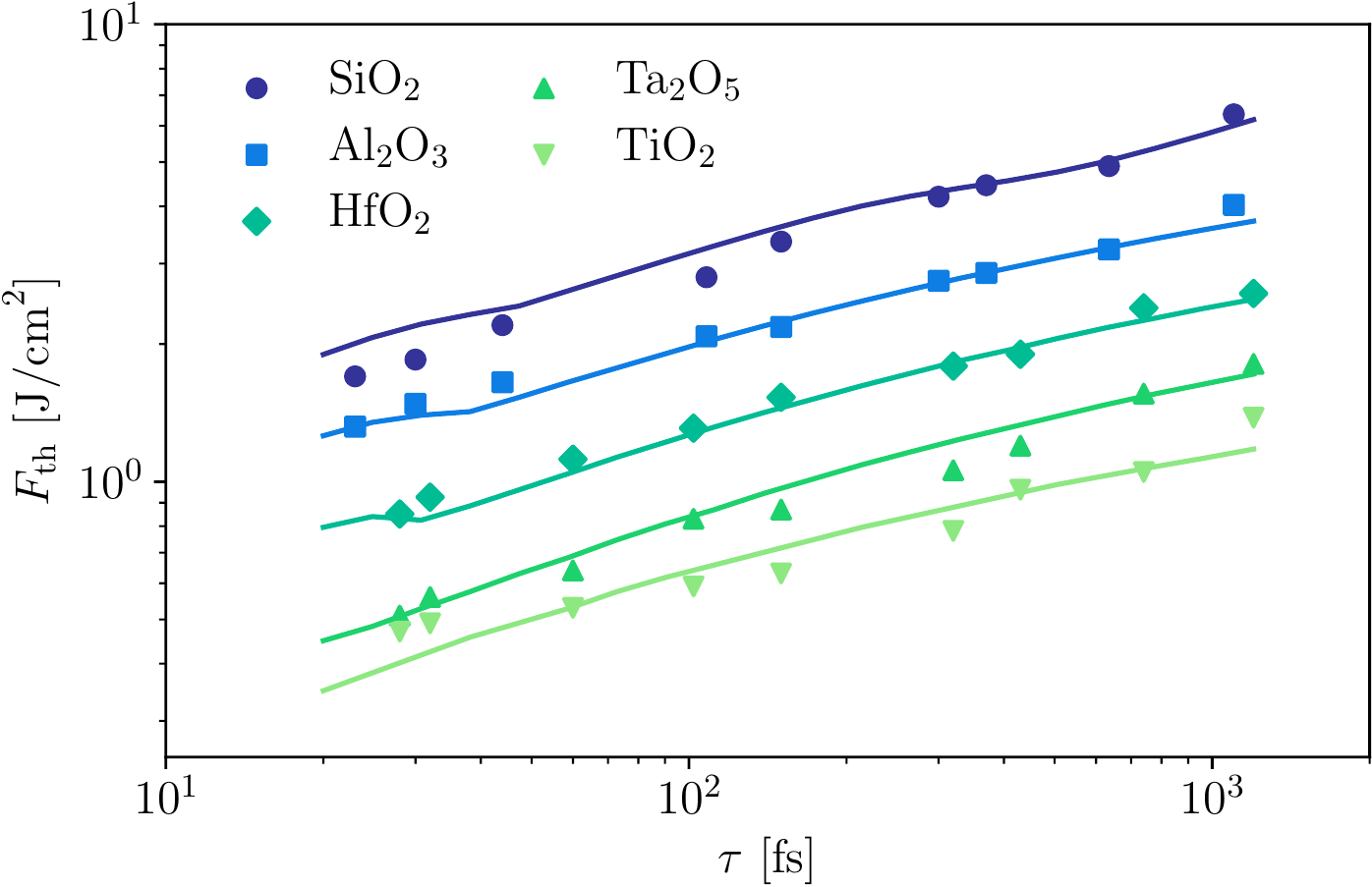}
\caption{(Color online) Comparison between DRE calculations (solid curves) and experimental measurements (shapes) of fluence thresholds as a function of pulse duration for various dielectric materials. The experimental data sets are from \cite{mero2005-2}. The parameters used for the DRE calculations are given in Table~\ref{tablePara}.}\label{figFthMat}
\end{figure}

The laser-induced damage threshold is a common reference to benchmark laser-induced dielectric breakdown models. It is often referred to as the minimum laser fluence $\Fth$ needed to cause permanent structural modifications to the material. On the plasma formation timescale, the laser-induced damage threshold is associated with the minimum laser fluence needed to create a plasma density $\rho \gtrsim \rho_c$ for which the medium becomes opaque to radiation with photon energy $\hbar\omega$. Based on the complex refractive index given at Eq.~\eqref{eq:refractive_index}, equating the real and imaginary parts gives the critical density that follows:
\begin{align}
\rho_c = \left(\frac{\epsilon_0 m_r}{q^2}\right) n_0^2\left(\omega^{2} + \gamma^{2}\right). \label{eq:rhoc}
\end{align}
To benchmark the DRE model, we have compared the results obtained by numerical integration of the underlying equations (see Sec.~\ref{sec:delayed_rate_model}) with the experimental data found in~\cite{mero2005-2}. Computations were done as in Sec.~\ref{sec:comparison} while scanning both the pulse duration $\tau$ and laser fluence $F$. When the maximum carrier density reached the critical density $\rho_c$ [see Eq.~\eqref{eq:rhoc}], the fluence is identified as the fluence threshold. Results are shown in Fig.~\ref{figFthMat}. Fit parameters are given in Table~\ref{tablePara}.

\begin{table}
\centering
\begin{tabular}{c c c c c c}
 	  							   & SiO$_2$ & Al$_2$O$_3$ & HfO$_2$ & Ta$_2$O$_5$ & TiO$_2$ \\ \hline
$n_0$ 							   & 1.45    & 1.76        & 2.09    & 2.1         & 2.52 \\
$\Eg$ [eV] 						   & 9.0     & 6.5         & 5.1     & 3.8         & 3.3\\
$\rhomol\,[10^{28}/\mathrm{m}^3]$  & 2.20    & 2.35        & 2.77    & 1.12        & 3.19 \\
$\sigmol\,[10^{-19}/\mathrm{m}^2]$ & 0.661   & 1.33        & 1.24    & 2.50        & 1.08 \\
$\gamma_r$ [ps$^{-1}$] 			   & 4.0     & 0.0    	   & 0.0     & 0.0         & 0.0\\\hline
$\gamma$ [fs$^{-1}$] 			   & 2.0     & 1.0         & 0.5     & 0.4         & 0.5\\
$m_e$ 						  	   & 1.0     & 0.8         & 0.4     & 0.5         & 0.3\\
$m_h$ 							   & 1.0     & 1.0         & 1.0     & 1.0         & 1.0\\
\hline
\end{tabular}
\caption{Dielectric material parameters associated with the DRE fits given in Fig.~\ref{figFthMat}. Typical values for the linear refractive index $n_0$, the bandgap $\Eg$, the recombination rate $\gamma_r$ and the molecular density $\rhomol$ are gathered from various references. To estimate the molecular cross-section $\sigmol$, we have summed the individual cross-sections of the constitutive atoms, calculated as the area of a circle with a radius equal to the covalent radius. The plasma damping rate $\gamma$ and the effective mass of the electrons $m_e$ (in units of the free electron mass $m_0$) are set by fitting experimental data for the pulse-length dependence of the fluence threshold (see Figs.~\ref{figFthMat} and~\ref{figFth}).}
\label{tablePara}
\end{table}

In  practice, the DRE computations shown in Fig.~\ref{figFthMat} rely only on two ``free'' parameters ($\gamma$ and $m_e$). To optimize the search for the best combination, we proceed as follows. First, we set the plasma damping rate $\gamma$ to adjust the overall scaling trend of the curve to obtain a reasonable agreement with a power-law fit of the experimental data (see below for details). Then, the effective mass parameter $m_e$ is chosen to fit the height of the corresponding data set. The parameters are not completely independent however [see, e.g., Eq.~\eqref{eq:rhoc}] and it is sometimes necessary to iterate the procedure for the final set of parameters. Nevertheless, the computed curves given in Fig.~\ref{figFthMat} show that DRE succeeds at reproducing the global trend of the experimental measurements over several orders of magnitude of both pulse duration and fluence threshold. Even better fits are obtained if more free parameters are used (e.g., $\gamma$, $m_e$, $m_h$, and $\Eg$).

We emphasize that effective bandgap and mass values are typically obtained by nondestructive measurement methods, where the sample integrity is only slightly perturbed. By definition, assessing the fluence threshold implies driving the material away from the ground state and potentially inducing significant changes to its band structure. The fit values should thus be interpreted with care. Note also that effective masses are usually tensors, to account for the anisotropy of the band structure. Simulations with DRE show that the mass parameters have a significant impact on the damage threshold, which in turn suggests that the orientation of the sample with respect to the laser polarization may play an important role. This effect is likely to be more pronounced in anisotropic crystalline structures. In particular, \textit{ab initio} calculations of the electronic band structure of $\mathrm{HfO}_2$ show that the effective masses along the different crystal planes can vary by more than an order of magnitude~\cite{Garcia2004}. The effective mass parameters given in Table~\ref{tablePara} are consistent with these calculations if they are considered as effective mass values averaged over the different crystal directions.

\begin{figure}
\centering \vspace{2mm}
\includegraphics[width=1.0\columnwidth]{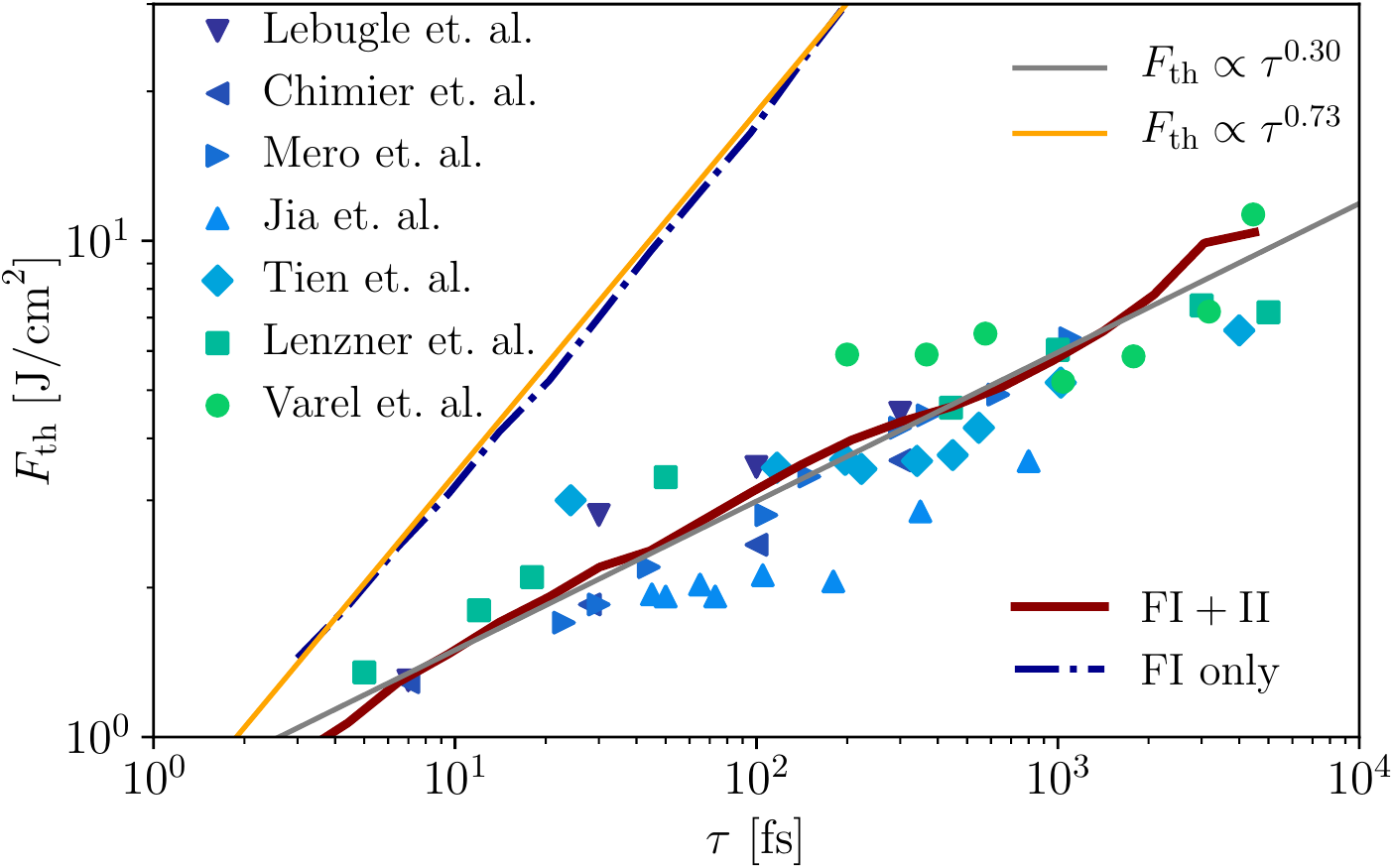}
\caption{(Color online) Comparison between the calculated and measured fluence thresholds as a function of pulse duration for fused silica. The experimental data sets are from \cite{lebugle2014,chimier2011,mero2005-2,jia2003,tien1999,lenzner1998,varel1996}. The red curve is calculated with DRE and the parameters from Table \ref{tablePara}. The blue dash-dotted curve was calculated with $\gib=0$.}\label{figFth}
\end{figure}

In Fig.~\ref{figFth}, we compare DRE with seven experimental data sets for fused silica. The typical trend across the experiments is that the fluence threshold follow a power-law dependence $\Fth \propto \tau^{\kappa}$, with $\kappa\simeq 0.3$ for $\tau< 10~\mathrm{ps}$. We could reproduce that trend using DRE and the parameters for SiO$_2$ in Table~\ref{tablePara}. Experimental data is lacking to rigorously test the model for pulse duration $<10$ fs. However, it is likely that DRE could be improved for such cases to include transient, field-cycle time scale process contributions (see Sec.~\ref{sec:discussion} for details). On the other hand, when neglecting laser heating (labelled as \textit{FI only}), which disables impact ionization completely, the scaling agreement is lost ($\kappa\simeq 0.73$). This supports the fact that near damage threshold impact ionization plays an important role in the dielectric breakdown process, even for few-femtosecond pulse duration.

\section{Discussion}\label{sec:discussion}

We have presented the DRE model as a potential replacement of MRE to study the plasma formation dynamics during laser-induced breakdown in dielectrics. Both models improve upon the SRE model by dealing with the time delay it takes for charge carriers to gain sufficient kinetic energy from the laser field to allow the creation of new charge carriers through impact ionization and trigger an ionization avalanche. DRE and MRE predict similar delays for the first impact ionization events to occur and for a potential ionization avalanche to unfold, with characteristic values for avalanche in the 80~fs range, in agreement with trusted experiments~\cite{stuart1995,tien1999}. Extended comparison of DRE predictions with experimental data for fused silica shows that the observed damage threshold scaling ($\Fth\propto \tau^{0.3}$) can only be explained if laser-heating of the charge carriers and subsequent carrier-impact ionization is taken into account. We have shown that DRE depends on a limited number of parameters that can be unambiguously associated with effective material properties.

There are a number of technical advantages for using DRE instead of MRE. In particular, DRE requires solving less equations, offering interesting possibilities for large scale, three-dimensional calculations where computational efficiency is important. Moreover, in the three-dimensional simulations of laser induced breakdown, e.g., using the finite-difference time-domain (FDTD) or the Particle-in-cell (PIC) frameworks, it is common to see high-contrast structures in the plasma density that strongly enhance or suppress the local electromagnetic field (see, e.g.,~\cite{Deziel2018}). This causes significant variations in the local ponderomotive energy of the charge carriers and, in turn, of the critical energy for impact ionization [see Eq.~\eqref{eq:E_critical}]. For MRE, this implies that numerical convergence is dictated by the number of rate equations used. This number must be chosen beforehand to account for the peak values of $\Ec$ over the entire simulation and throughout the material domain. This is an important drawback for MRE that should not be overlooked. For DRE, defined by a closed set of equations, this is not an issue.

Finally, it is important to recall that rate equation models in general describe laser-induced breakdown at the field-cycle-averaged level. Future improvements should include proper treatment of photon-assisted avalanche, often referred to as cold ionization avalanche (see, e.g.,~\cite{Rajeev2009}), as well as potential sub-cycle process contributions (see, e.g.,~\cite{macdonald2017,zhokhov2014}). 

\section{Conclusion}\label{sec:conclusion}

We have provided a theoretical framework to study plasma formation during femtosecond laser-induced breakdown in dielectrics on a field-cycle average, statistical level. The model improves upon the current approaches by providing an explicit, closed-formed treatment of the charge-carrier laser-heating process that precedes the onset of carrier-impact ionization and a potential collisional ionization avalanche. In particular, we have shown that the model we propose can reproduce damage-threshold data over several orders of magnitude in both the laser pulse duration and laser fluence, while relying on a limited number of parameters related to effective material properties. A side benefit of the model is its computational efficiency that opens possibilities for large-scale, three-dimensional modelling of laser-induced breakdown and structural pattern formation in transparent media.

\appendix

\section{Definitions for the mass symbols used in this paper}\label{appendix:masses}

In this paper, the effective mass of the electrons in the conduction band (CB) is denoted by $m_e$ and the effective mass of the holes in the valence band (VB) is $m_h$. In some cases, the reduced mass $m_r^{-1}=m_e^{-1} + m_h^{-1}$ is used. For example, the IBH rate $\gib^{e}$ for electrons is calculated with $m_e$ and the IBH rate $\gib^{h}$ for holes is calculated with $m_h$ to give a total IBH rate $\gib=\gib^{e}+\gib^{h}$, which can be calculated with $m_r$. The total ponderomotive energy of electrons and holes (see Appendix~\ref{sec:laser_currents}) is also calculated with $m_r$. Finally, we refer to the free electron mass with the symbol $m_0$.

\section{Drude description of the laser-plasma dynamics} \label{sec:laser_currents}

The instantaneous current $\tilde{i}(t)$ associated with the motion of a charge carrier (electron or hole) is conveniently described at a statistical-continuum level by the Drude-like single-carrier model that follows:
\begin{align} 
\frac{\mathrm{d}\tilde{i}(t)}{\mathrm{d}t} = -\gamma \tilde{i}(t) + \frac{q^2}{m} \tilde{E}(t),
\end{align}
where $\tilde{E}(t)$ is the electric field of the laser. Parameters $q$ and $m$ are the charge and mass of the charge carrier, respectively. Collisions are included phenomenologically \textit{via} the damping rate $\gamma$. Given the carrier density $\rho(t)$, a current density is then defined as $\tilde{J}(t) =  \rho(t)\tilde{i}(t)$.

For $\tilde{E}(t)=E\cos(\omega t)$, the steady-state solution for the single-carrier current is:
\begin{align} 
\tilde{i}(t) = \frac{q^2 E}{m(\gamma^2 + \omega^2)} \left[ \omega\sin(\omega t) + \gamma\cos(\omega t) \right].
\end{align}
Then, the power transferred instantaneously from the laser field to the charge carrier is given by 
\begin{align} 
\tilde{P}(t) &= \tilde{i}(t) \cdot \tilde{E}(t), \nonumber \\
&= \frac{q^2 E^2}{m(\gamma^2 + \omega^2)} \left[ \omega\sin(\omega t)\cos(\omega t) + \gamma\cos^2(\omega t) \right]. \label{eq:power}
\end{align}
The two terms in the square brackets are associated with the ponderomotive energy and inverse bremsstrahlung heating, described below.

\subsection{Ponderomotive energy} The first term in the square brackets of Eq.~\eqref{eq:power} represents a carrier that gains a certain amount of energy during half of an optical cycle, before losing it during the other half, resulting in no net energy gain or loss. This is often referred to as the ponderomotive energy, whose instantaneous expression is given by the integral of the ponderomotive power, i.e., of the first term in Eq.~\eqref{eq:power}, such that
\begin{align} 
\tilde{\mathcal{E}}_p(t) &= \int q^2 E^2 \frac{\omega\sin(\omega t)\cos(\omega t)}{m(\gamma^2 + \omega^2)} \mathrm{d}t, \nonumber \\
&= \frac{q^2|\tilde{E}(t)|^2}{2m(\gamma^2 + \omega^2)}.
\end{align}
In general, the ponderomotive energy is expressed instead in terms of its cycle-averaged expression
\begin{align}
\mathcal{E}_p = \langle \tilde{\mathcal{E}}_p(t) \rangle = \frac{q^2E^2}{4m(\gamma^2 + \omega^2)}. \label{eq:ep}
\end{align}
that reduces to the usual, free-particle expression $\mathcal{E}_p = q^2E^2/4m\omega^2$ in the limit where $\gamma = 0$. 

\begin{figure}
	\centering\vspace{2mm}
	\includegraphics[width=1.0\columnwidth]{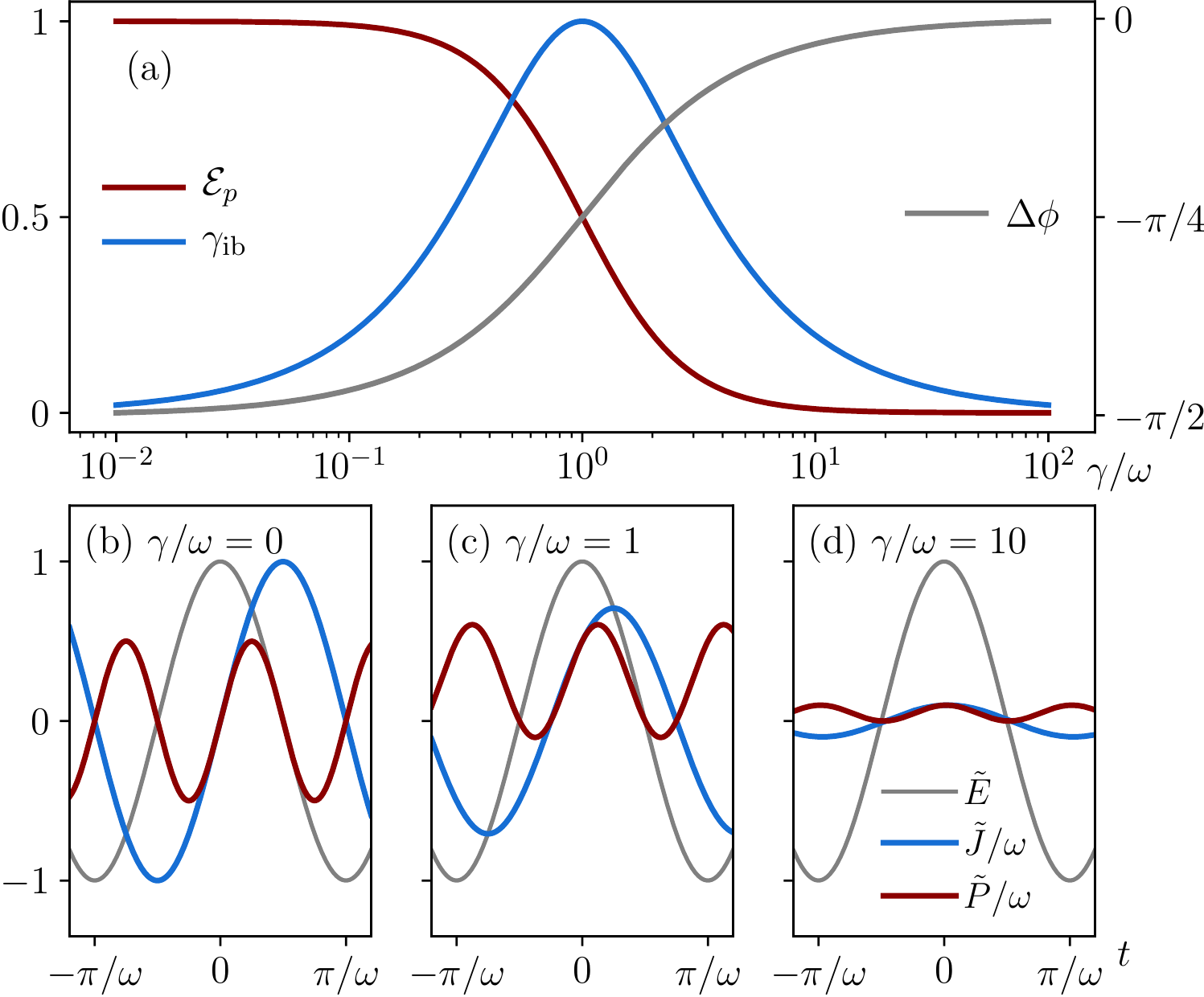}
	\caption{(Color online) Physical insight into the continuum expressions obtained with the Drude model. In (a), as a function of the plasma damping rate $\gamma$, the normalized ponderomotive energy [Eq.~\eqref{eq:ep}] and laser-heating rate [Eq.~\eqref{eq:gib}] (both refer to the y axis on the left), as well as the phase shift between current density and field oscillations [$\Delta \phi = \arctan(-\omega/\gamma)$] ($\Delta \phi$ refers to the y axis on the right). In (b) to (d), normalized comparison of the temporal evolution of the electric field, current density, and power for three values of damping.}\label{figDrude}
\end{figure}

\subsection{Inverse bremsstrahlung heating} The last term of Eq.~\eqref{eq:power} is associated with the absorption by the charge carrier of electrical power from the laser field resulting in a net energy gain after each optical cycle. The rate at which a quantum of light is absorbed is obtained by dividing the last term of Eq.~\eqref{eq:power} by the energy of a photon $\hbar\omega$, thus defining an instantaneous laser-heating rate as
\begin{align} 
\tilde{\gamma}_\mathrm{ib}(t) = \frac{\gamma}{\hbar\omega}\frac{q^2 |\tilde{E}(t)|^2}{m(\gamma^2 + \omega^2)}=\frac{2\gamma}{\hbar\omega}\tilde{\mathcal{E}}_p(t).
\end{align}
When averaged over a field cycle:
\begin{align} 
\gib = \langle \tilde{\gamma}_\mathrm{ib}(t) \rangle = \frac{\gamma}{\hbar\omega}\frac{q^2 E^2}{2m(\gamma^2 + \omega^2)}=\frac{2\gamma}{\hbar\omega}\mathcal{E}_p.\label{eq:gib}
\end{align}

Physical insight into the continuum model for the ponderomotive energy and laser-heating rate in the presence of collisions is provided in Fig.~\ref{figDrude} [where we used a constant value for $\rho(t)$]. In the free-particle limit ($\gamma = 0$), no photon is absorbed, which results in a purely ponderomotive regime ($\gamma_\mathrm{ib} = 0$). But as $\gamma$ is increased, the amplitude of the current density decreases and the phase difference with respect to the field oscillations $\Delta \phi = \arctan(-\omega/\gamma)$ gradually shifts from $-\pi/2$ to $0$, with no energy transfer to the charge carriers in the limit $\gamma \rightarrow \infty$. Optimal heating occurs when $\gamma = \omega$.

\section{Keldysh model for field ionization in solid-state dielectrics} \label{sec:fi}

\begin{figure}
\centering\vspace{2mm}
\includegraphics[width=1.0\columnwidth]{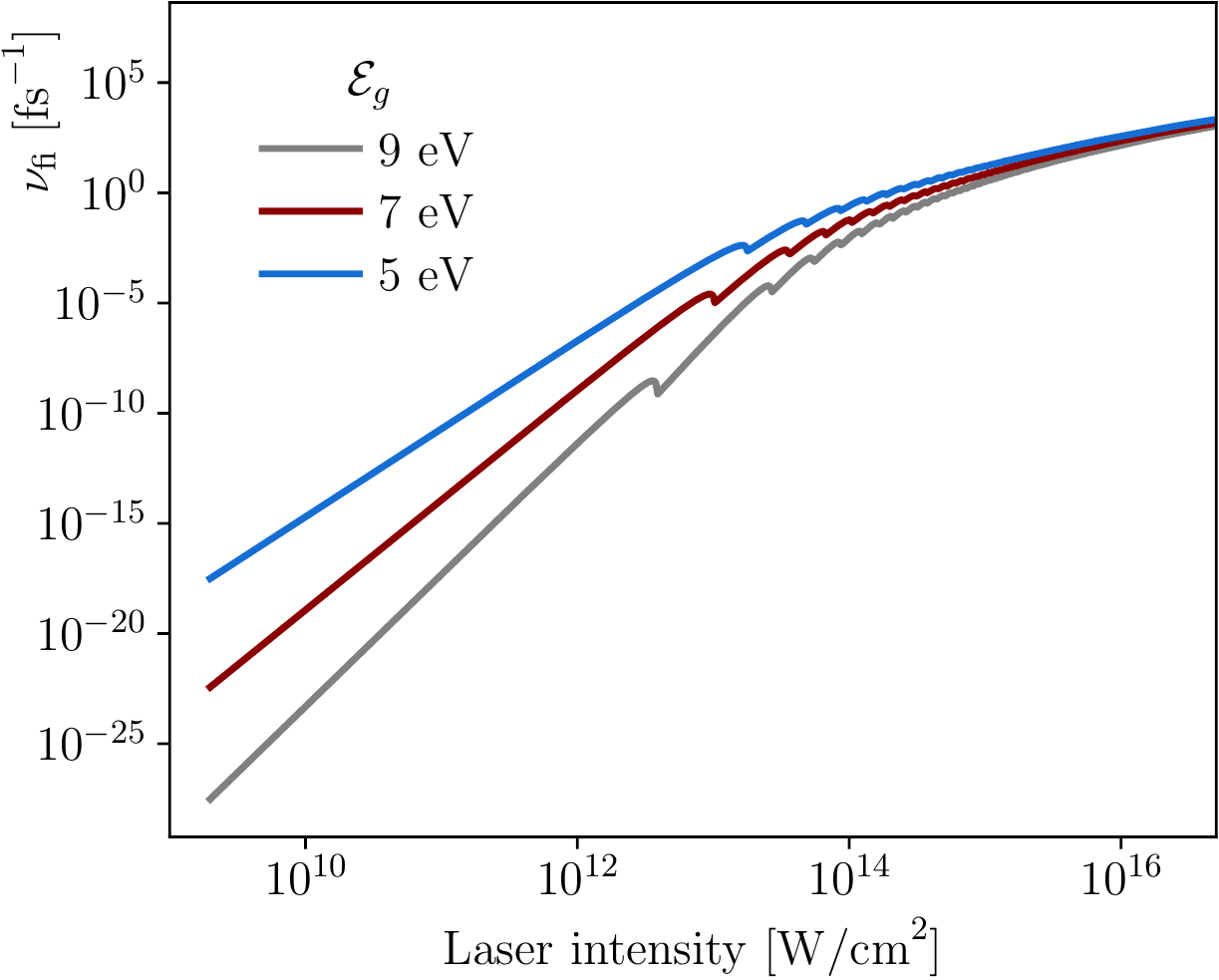}
\caption{(Color online) FI rates predicted by Keldysh [see Eq.~\eqref{eq:W_Keldysh}] for different values of the band gap energy $\Eg$. Parameters are $\lambda=800$ nm, $m_e=m_h=m_0$, $\rhomol=2\cdot 10^{28}$ m$^{-3}$ and $n_0=1.5$.}\label{figFi}
\end{figure}

The production rate of electron-hole pairs (in $\mathrm{m}^{-3}\mathrm{s}^{-1}$) induced by a strong laser field $\tilde{E}(t)=E\cos(\omega t)$ in a solid-state dielectric with bandgap energy $\mathcal{E}_g$ is given by the Keldysh relation (for details, see ref.~\citep{balling2013}, Sec. 2.3.1 and \cite{couairon2007}, Sec.~2.3)
\begin{align}\label{eq:W_Keldysh}
W = \frac{4\omega}{9\pi}\left(\frac{m_r\omega}{\hbar\sqrt{x_1}}\right)^{3/2}\left(\frac{\pi}{2\mathcal{K}(x_2)}\right)^{1/2}\sum_{n=0}^{\infty}e^{-(k+n) \alpha}\,\Phi\left(x_3\right)
\end{align}
where
\begin{align}
x_1=\frac{\Gamma^2}{1+\Gamma^2}; \,\,\, x_2 = \frac{1}{1+\Gamma^2}&; \,\,\, x_3=\sqrt{\beta\left(2\nu+n\right)}\\
\alpha = \pi\frac{\mathcal{K}(x_1) - \mathcal{E}(x_1)}{\mathcal{E}(x_2)}; \,\,\, \beta=&\frac{\pi^2}{2\mathcal{K}(x_2)\mathcal{E}(x_2)}; \,\,\, \nu = k-x\\
\Gamma = \sqrt{\frac{\mathcal{E}_g}{2\mathcal{E}_p}}; \quad x=\frac{2}{\pi}\frac{\mathcal{E}(x_2)}{\sqrt{x_1}}&\frac{\mathcal{E}_g}{\hbar\omega} ; \quad k = \lfloor x + 1\rfloor
\end{align}
with $\mathcal{K}()$ and $\mathcal{E}()$ being the complete elliptic integrals of the first and second kind, respectively, $\Phi()$ being Dawson's integral, and $\lfloor\ldots\rfloor$ denoting the integral part of the argument. The free-particle ponderomotive energy $\mathcal{E}_p = q^2E^2/2m_r\omega^2$ defines the Keldysh parameter as
$\Gamma = (\omega/qE)\sqrt{m_r\mathcal{E}_g}$ (see ref.~\citep{balling2013}).

To get an FI rate $\nufi(\gamma)$ compatible with rate-equation models [e.g., Eqs.~\eqref{eq:sre},~\eqref{eq:mre2}, and~\eqref{eq:dre}], the Keldysh rate $W$ (in $\mathrm{m}^{-3}\mathrm{s}^{-1}$) is divided by the molecular density $\rho_\mathrm{mol}$ (in $\mathrm{m}^{-3}$) of the material. The resulting, single-molecule ionization rate $\nufi = W/\rho_\mathrm{mol}$ is plotted in Fig.~\ref{figFi} for different values of the band gap energy $\Eg$.

\subsection*{Acknowledments}
C.V. acknowledges financial support from the Natural Sciences and Engineering Research Council of Canada (NSERC) through the College and Community Innovation Program - Innovation Enhancement Grants (CCIPE 517932-17) and the Fonds de
	recherche du Qu\'ebec - Nature et technologies (FRQNT) through the Programme de recherche
	pour les chercheurs et les chercheuses de coll\`ege (2019-CO-254385).
%

\end{document}